\documentclass[singlecolumn,showpacs,amsmath,amssymb,amsfonts,nofootinbib,plb]{revtex4}
\usepackage{graphicx}
\usepackage{dcolumn}
\usepackage{bm}
\usepackage{ulem}
\usepackage{overpic}
\usepackage{amssymb}

\newcommand{\nn}{\nonumber}

\def\beq{\begin{equation}}
\def\eeq{\end{equation}}

\usepackage{soul}
\usepackage{cancel}

\usepackage[usenames]{color}

\begin{document}
\input{epsf}

\title{Infrared renormalon  effects in color dipole TMD PDF  at small-$x$}

\author{Nahid Vasim}
\email{nvasim@myamu.ac.in}
\author {Raktim Abir}   
\email{raktim.ph@amu.ac.in}

\affiliation{Department of Physics, Aligarh Muslim University, Aligarh - $202002$, India.}

 \begin{abstract}
The uncertainties from the infrared renormalons in the (color dipole) gluon distribution is estimated. It is shown that non-linear saturation effects at small-$x$ shift the first IR pole at the Borel plane from $2/\beta_2$ to  $1/\beta_2$. As a consequence  the estimated uncertainty is found to be ${\cal O}\left(\Lambda_{\rm QCD}^2\right)$ instead of ${\cal O}\left(\Lambda_{\rm QCD}^4\right)$. 
 \end{abstract}


\date{\today}
\maketitle

\section{Introduction}

Perturbation series in quantum field theory are usually divergent even after term by term renormalization of mass and charge. 
This was first argued by Freeman Dyson in early fifties on the eve of developement of quantum electrodynamics (QED) \cite{Dyson:1952tj}. 
In case of QED it was shown that individual terms of the perturbation series 
$
  \sum_{n} {\cal C}_n \alpha_s^n
$
first decrease and gradually approaches to a minimum and then start  to increase without any limit. The index of the minimum term estimated to be of the order $137$. 
The ever increasing terms make the perturbation series divergent.  
However fortunately this no way restricts predictions from the perturbation series for practical applications. 
Nevertheless this anomaly raises important questions on very foundation upon which the theory is built. 
 B. Lautrap was first who point out that there could be single gauge invariant diagram that contribute to amplitude to grow it like $n!$ in the $n$-th order term of the perturbation series \cite{Lautrup:1977hs} making it a divergent series. 
Not just in QED, this subtlety also translates to QCD, where strong coupling acts as the expansion parameter of the series.  
Some non-perturbative effects in QCD are actually stems from the asymptotic nature of the perturbation series and corroborated by the fact that the coefficients ${\cal C}_n$ often have factorial-growth with $n$. 
There are resummation procedure to regularize divergence series $e.g.$  Borel resummation. 
For factorially divergent series Borel summation is mostly used.  
The source of this divergence of perturbation series in QCD is then reflected through the presence of pole singularities \cite{Mueller:1984vh} on the real axis of the Borel plane leading to (a) Infrared Renormalons (IR), (b) Ultraviolet Renormalons (UR) and the (c) Instantons \cite{Beneke:1998ui}.  
If the Borel integral has no singularity in the positive real axis and the terms in the series do not increase faster than the factorial growth, the divergent series is Borel summable.  In QCD infrared renormalons are singularities on the positive $b$-axis due to the integration over infrared regions and presence of them spoil the Borel summability of the series. 
In fact  it was argued that this Borel non-summable contribution predicts a specific high-energy phenomenon which calls for a  growth of multi-jet production with high multiplicity. Phenomenologically, the phenomena leads to apparently anomalous events that would look as highly isotropic multi-jet events \cite{Zakharov:1992bx}. 
\\
 
 \noindent QCD renormalons in small-$x$ was first addressed by Levin  \cite{Levin:1994di}. In QCD high energy scattering events generally involve cascade of gluons. This is because unlike photon  primary gluons with high virtuality themselves  emit further secondary gluons in high energy. 
At high enough energy this rapidly growing cascade of gluons grow to an extent that fusion of multiple gluons to single gluon begin. Over the time this fusion of gluons develops a statistical detailed balance with the reverse phenomena of gluon cascade. 
This detail balance of gluon number density leads to the origin of the phenomena of gluon saturation with the emergence of a dynamically generated and energy dependent  momentum scale $Q_s$ known as saturation scale. 
Above  this saturation scale one may safely assumes independent stochastic multiple scattering approximations but below this scale this is no longer valid.  Highly correlated non-linear gluon interactions prevail over normally distributed random interactions below the saturation scale.  
\noindent  In the small momenta regions, within the window of momentum scale $\Lambda_{\rm QCD}$ and saturation scale $Q_s$ where the running coupling becomes large, infrared renormalons is believed to be the source of the divergence in the perturbation series and gives estimates to the uncertainty due to non-perturbative effects. 
In this article we revisit the estimation of the uncertainties from the infrared renormalons in the (color dipole) gluon distribution employing our recently derived expression that is valid in small transverse momentum region \cite{Abir:2018hvk}.  
We have shown that non-linear saturation effects at small-$x$ shift the first IR pole at the Borel plane towards zero from $2/\beta_2$ to  $1/\beta_2$ where $\beta_2$ is beta function of QCD.  This leads to enhanced non-perturbative uncertainty ${\cal O}\left(\Lambda_{\rm QCD}^2\right)$ instead of ${\cal O}\left(\Lambda_{\rm QCD}^4\right)$ for the (color dipole) gluon distribution.

 \section{Infrared renormalon in (color dipole) gluon distribution}

\noindent We start with the relation between dipole amplitude ${\cal N}$ and the unintegrated dipole gluon distribution  ${\cal F}$ \cite{Kovchegov_book}
\begin{eqnarray}
\int d^2b~{\cal N}(r_\perp,b_\perp,x) = \frac{2\pi}{N_c} \int d^2k_\perp \left(1-e^{ik_\perp.r_\perp}\right)\alpha_s(k_\perp^2)\frac{1}{k_\perp^2} {\cal F} \left(x, k_\perp\right),
\label{1}
\end{eqnarray}
Now at or around  saturation scale dipole distribution ${\cal F}$ often taken to be  a $Gaussian \otimes quadratic$ distribution of transverse momentum 
\begin{eqnarray}
 {\cal F}(x,k_\perp)& \propto &  ~\frac{k_\perp^2}{Q_{s}^2(x)}~
 \exp\left[{-\frac{k_\perp^2}{Q_{s}^2(x)}}\right],  \label{B}  \\ 
 &\approx & \frac{k_\perp^2}{Q_{s}^2(x)}    ~~~~~ ({\rm when}~k_\perp^2 \ll Q_s^2(x)).  \label{C}
\end{eqnarray} 
Eq.\eqref{B} can be readily derived from the dipole $S$-matrix that is Gaussian ($e.g$ in GGM, MV or in phenomenological GBW model) in scaling variable $t$ $(\equiv r_\perp Q_s)$
\begin{eqnarray}
S(r_\perp,x) = \exp \left(-\kappa r_\perp^2 Q_s^2(x)\right)
\end{eqnarray}
where $\kappa \approx 1/4$ is a constant. Strictly speaking  this is valid around the saturation region and not deep inside  where multiple scatterings are highly correlated and can't be taken just as collection of random independent scatterings. Nonetheless the expression often taken to extrapolate it in the low momentum limit and leads to the conclusion that the unintegrated dipole gluon distribution behaves as $k_\perp^2/Q_s^2(x)$ in the low momentum. 
The relation in Eq.\eqref{1} between dipole amplitude  and the unintegrated dipole gluon distribution follows from the two-gluon exchange. While it may not be strictly valid in the highly nonlinear regime nonetheless we will take Eq.\eqref{C} to estimate the contribution to the dipole amplitude from the saturation region with $k_\perp < Q_s(x)$. 
In the kinematic region where the gluon distribution varies as $\phi\propto k_\perp^2/Q_s^2$, the contribution to dipole amplitude from the saturation region is proportional to 
  \begin{eqnarray}
   \int \frac{d^2k_\perp}{k_\perp^2} \left(1-e^{ik_\perp.x_\perp}\right)\alpha_s(k_\perp^2)\frac{ k_\perp^2}{Q_s^2}~. 
 \label{E}
  \end{eqnarray}
We now introduce the renormalization scale $\mu$ writing,
   \begin{eqnarray}
 \alpha_s \left(k_\perp^2\right)=\frac{\alpha(\mu^2)}{1+\alpha(\mu^2)\beta_2 \ln \left(k_\perp^2/\mu^2\right)}.
   \end{eqnarray}
We substitute this in Eq.\eqref{E} and expand in power of $\alpha_s(\mu)$, to obtain, 
\begin{eqnarray}
\frac{r_\perp^2}{Q_s^2}\alpha_s(\mu) \sum_{n=0}^{\infty} \left(-\alpha(\mu)\beta_2\right)^2 \int_0^{Q_s^2}dk_\perp^2 k_\perp^2 \ln^n\frac{k_\perp^2}{\mu^2}
 \end{eqnarray}
Define $\zeta\equiv \ln \left(\mu^2/k_\perp^2\right)$ we rewrite
\begin{eqnarray}
\frac{r_\perp^2}{Q_s^2} \mu^4 \alpha_s(\mu^2) \sum_{n=0}^{\infty} \left(\alpha(\mu)\beta_2\right)^2 \int_{\ln\left(\mu^2/Q_s^2\right)}^{\infty}d\zeta~\zeta^n~e^{-2\zeta} 
 \end{eqnarray}
The expression contains an incomplete gamma integral for $\zeta$ and results in a diverging series
  \begin{eqnarray}
  \frac{r_{\perp}^2}{2Q_s^2}\mu^4\alpha\left({\mu}\right)
    \sum_{n =0}^{\infty}
   \left(\frac{\alpha(\mu) \beta_2}{2} \right)^n~n!~{\cal C}
   \label{facrise} 
  \end{eqnarray}
  where ${\cal C}$ is just a correction factor mimics the deviation from complete gamma function, 
    \begin{eqnarray}
  {\cal C} = 1-\frac{Q_s}{\mu} \sum_{k=n+1}^{\infty} \frac{1}{k!}\ln^k \frac{\mu^2}{Q_s^2}.
  \label{F}
    \end{eqnarray}
This clearly shows for large enough $n$ the integral is dominated  by upper limit so that its lower limit can be set equal to zero or otherwise ${\cal C}$ can be approximate to unity leaving only the rising factor proportional to $n!$. This is the typical effect of infrared QCD renormalons. Divergent series with rising factorial in the coefficients are  best treated using Borel summation techniques. Namely we rewrite the series as 
    \begin{eqnarray} 
 -\frac{2}{\beta_2}\frac{r_\perp^2}{2Q_s^2}\mu^4 \int_{0}^{\infty}db~e^{-b/\alpha(\mu^2)} \frac{1}{b-2/\beta_2}  
   \end{eqnarray}
 The pole at $b=2/\beta_2$ is known as the IR renormalon pole in the complex $b$-plane. The $b$-integral is divergent as  the  renormalon pole is on the positive real axis  and hence the series is not Borel resummable. 
 At best one may estimate the size of the IR renormalon uncertainty by simply taking the residue of the renormalon pole, which gives, 
    \begin{eqnarray} 
  \sim \frac{r_\perp^2}{Q_s^2} \mu^4 \exp\left(-\frac{2}{\alpha_s(\mu^2)\beta_2}\right)
    \end{eqnarray}
or in terms of $\Lambda_{QCD}$, 
\begin{eqnarray}
 \sim\frac{r_{\perp}^2}{Q_s^2}\mu^4 \exp\left(-\frac{2}{\alpha_s(\mu^2)\beta_2}\right)=\frac{r_{\perp}^2}{Q_s^2}\Lambda^4_{QCD}
 \label{K}
\end{eqnarray}
Non-perturbative origin of the uncertainty reflects from the fact that the result is proportional to $\Lambda_{QCD}^4$.  \\

 \section{Infrared renormalon revisited}
 
\noindent  Recently we derive an analytical expression for unintegrated color dipole gluon distribution  at asymptotically  small transverse momentum. As small transverse momentum corresponds to large transverse separation, the new expression is derived by performing the Fourier transform of $S$-matrix for large dipoles with large transverse separations. In particular we take Levin-Tuchin solution of Balitsky-Kovchegov  equation to get the results. Though the result found out to be in the form of a series of Bells polynomials, when resumming in leading log accuracy, showing up Sudakov like soft factor \cite{Abir:2018hvk}
\begin{eqnarray}
 &&{\cal F}(x,k_\perp) \propto 
\ln\left(\frac{k_\perp^2}{4Q_s^2}\right)\exp\left[-\tau \ln^2\left(\frac{k_\perp^2}{4Q_s^2}\right)\right]
\label{G}
\end{eqnarray}
where $\tau \approx 0.2$ is a constant.
Its interesting to note that the small-$x$ evolution kills the $k_\perp^2/Q_s^2$ behaviour from Eq.\eqref{B} to Eq.\eqref{C} in the small momentum limit and modifies it from $linear\otimes normal$ to $logarithmic \otimes
 log~ normal$  as given in Eq.\eqref{G}. In the following we take Eq.\eqref{G} to estimate the associated uncertainties stemming from IR renormalon to have a comparative study with $e.g.$ Eq.\eqref{K}. 
We start by writing 
\begin{eqnarray}
 \int^{Q_s^2}\frac{d^2k_\perp}{k_\perp^2}\left(1-e^{ik_\perp.r_\perp}\right)\alpha_s(k_\perp^2)
    ~\ln\left(\frac{k_\perp^2}{4Q_s^2}\right)\exp\left[{-\tau \ln^2\left(\frac{k_\perp^2}{4Q_s^2}\right)}\right]
\end{eqnarray}
We expand the exponentials (with $k_\perp x_\perp\ll1$) and performing angular integration to get
\begin{eqnarray}
 && r_\perp^2\int^{Q_s^2}{dk_\perp^2}~\alpha_s(k_\perp^2)
~~\ln\left(\frac{k_\perp^2}{4Q_s^2}\right) \sum_{m=0}^{\infty}{\frac{(-\tau)^m}{m!}\ln^{2m}
\left(\frac{k_\perp^2}{4Q_s^2}\right)}
\end{eqnarray}
We now simplify by expressing the $m$-th term as $(2m+1)$-th derivative of dummy variable $\epsilon$ in the limit $\epsilon \rightarrow 0$,
\begin{eqnarray}
r_\perp^2\sum_{m=0}^{\infty}{\frac{(-\tau)^m}{m!}}\int^{Q_s^2}{dk_\perp^2}~\alpha_s(k_\perp^2)
~\lim_{\epsilon\rightarrow0}\frac{\partial^{2m+1}}{\partial\epsilon^{2m+1}}
{\left(\frac{k_\perp^2}{4Q_s^2}\right)^{\epsilon}}
\end{eqnarray}
Substituting the effective QCD coupling,
\begin{eqnarray}
 \alpha_s(k_\perp^2)=\frac{\alpha\left({\mu^2}\right)}{1+\alpha(\mu^2) \beta_2 \ln(k_\perp^2/\mu^2)}
\end{eqnarray}
as an expansion in powers of $\alpha\left(\mu^2\right)$
 \begin{eqnarray}
 &&  r_\perp^2\sum_{m=0}^{\infty}{\frac{(-\tau)^m}{m!}}\lim_{\epsilon\rightarrow0}
 \frac{\partial^{2m+1}}{\partial\epsilon^{2m+1}}{\left(\frac{\mu^2}{4Q_s^2}\right)^{\epsilon}}
 \sum_{n=0}^{\infty}(-1)^n\alpha^{n+1}(\mu)\beta^n_2 
\int^{Q_s^2}{dk_\perp^2}~\ln^n
  \left(\frac{k_\perp^2}{\mu^2}\right)
{\left(\frac{k_\perp^2}{\mu^2}\right)^{\epsilon}}\label{app1}
  \end{eqnarray}
~\\
We define $\xi=\ln\left( \mu^2/k_\perp^2\right)$ to perform the $k_\perp$-integration in Eq.\eqref{app1}:
\begin{eqnarray}
&&  \int^{Q_s^2}{dk_\perp^2}~\ln^n
  \left(\frac{k_\perp^2}{\mu^2}\right)
{\left(\frac{k_\perp^2}{\mu^2}\right)^{\epsilon}}\nn \\
&=&-\mu^2 \int^{\ln\left({\mu^2}/{Q_s^2}\right)}_{\infty}{d\xi}~e^{-\xi}(-\xi)^ne^{-\xi\epsilon}
  \nn\\
&=&(-1)^n \mu^2 \int_{\ln\left({\mu^2}/{Q_s^2}\right)}^{\infty}{d\xi}~\xi^n~e^{-\xi(1+\epsilon)}
  \nn  \\
  &=&(-1)^n  \mu^2 {\left(\frac{1}{1+\epsilon}\right)}^{n+1}\Gamma({n+1})\nn
  \end{eqnarray}
which yields a divergent series
\begin{eqnarray}
  &&  r_\perp^2\sum_{m=0}^{\infty}{\frac{(-\tau)^m}{m!}}\lim_{\epsilon\rightarrow0}
  \frac{\partial^{2m+1}}{\partial\epsilon^{2m+1}}{\left(\frac{\mu^2}{4Q_s^2}\right)^{\epsilon}}
  \mu^2\sum_{n=0}^{\infty}{\beta^n_2 }~\left(
\frac{{\alpha(\mu)}}{1+\epsilon}\right)^{n+1}n!
\end{eqnarray}
with rising factorials in the coefficients, similar to Eq.\eqref{facrise}, due to the IR renormalons. 
The inner series can be formally rewritten as
 \begin{eqnarray}
 \sum_{n=0}^{\infty}{\beta^n_2 }~
\left(
\frac{{\alpha(\mu)}}{1+\epsilon}\right)^{n+1}n!=\sum_{n=0}^{\infty} 
\beta_2^n \int_{0}^{\infty} db
~b^n~ 
\exp\left(-\frac{(1+\epsilon)~b}{\alpha(\mu)}\right)  
 \end{eqnarray}
where the integration variable $b$ can be taken as a complex dummy parameter over which the integration is to be performed in the complex Borel $b$-plane. We now introduce Borel summation technique to evaluate the factorial rising expansion by formally switching the integration and summation
\begin{eqnarray}
\sum_{n=0}^{\infty} 
\beta_2^n \int_{0}^{\infty} db
~b^n~ 
\exp\left(-\frac{(1+\epsilon)~b}{\alpha(\mu)}\right)  
\rightarrow \int_{0}^{\infty} db  ~\exp\left(-\frac{(1+\epsilon)~b}{\alpha(\mu)}\right)  \sum_{n=0}^{\infty} 
\beta_2^n b^n
\end{eqnarray}
Interestingly now the pole of the integral is shifted to $b=1/\beta_2$
unlike the previous case where it was at  $b=2/\beta_2$
\begin{eqnarray}
\int_{0}^{\infty} db  ~\exp\left(-\frac{(1+\epsilon)~b}{\alpha(\mu)}\right)  \sum_{n=0}^{\infty} 
\beta_2^n b^n
= \int_{0}^{\infty} db  ~\exp\left(-\frac{(1+\epsilon)~b}{\alpha(\mu)}\right) \frac{1}{1-\beta_2 b} \label{app2}
\end{eqnarray}
 Taking the residue the $b$-integration in Eq.\eqref{app2} found to be
 \begin{eqnarray}
   &\sim&-\frac{1}{\beta_2} \exp\left(-\frac{1+\epsilon}{\beta_2 ~\alpha(\mu)}\right)   \nn
 \end{eqnarray}
 The overall measure of the uncertainty from the IR pole at $1/\beta_2$ is then
 \begin{eqnarray}
 &\sim& -\frac{x_\perp^2 \mu^2}{\beta_2}\sum_{m=0}^{\infty}{\frac{(-\tau)^m}{m!}}
\lim_{\epsilon\rightarrow0}\frac{\partial^{2m+1}}{\partial\epsilon^{2m+1}}{\left(\frac{\mu^2}{4Q_s^2}\right)^{\epsilon}} ~
\exp\left(-\frac{1+\epsilon}{\beta_2 ~\alpha(\mu)}\right)  
\end{eqnarray}
Next we perform $2m+1$ derivative taking 
\begin{eqnarray}
 t&=&{\left(\frac{\mu^2}{4Q_s^2}\right)^{\epsilon}} ~
\exp\left(-\frac{1+\epsilon}{\beta_2 \alpha(\mu)}\right) \nn\\
\ln t &=&\epsilon\ln\frac{\mu^2}{4Q_s^2}-\frac{1+\epsilon}{\beta_2\alpha\left(\mu\right)}\nn\\
\frac{dt}{d\epsilon}&=&t\left(\ln\frac{\mu^2}{4Q_s^2}-\frac{1}{\beta_2\alpha(\mu)}\right)\nn\\
\frac{d^2t}{d\epsilon^2}&=&t\left(\ln\frac{\mu^2}{4Q_s^2}-\frac{1}{\beta_2\alpha(\mu)}\right)^2\nn\\
\Rightarrow\frac{d^{2m+1}t}{d\epsilon^{2m+1}}&=&t\left(\ln\frac{\mu^2}{4Q_s^2}-\frac{1}{\beta_2\alpha(\mu)}\right)^{2m+1}\nn
\end{eqnarray}
After taking the limit $\epsilon\rightarrow0$, the uncertainty is
\begin{eqnarray}
&\sim& r_\perp^2 \mu^2~\exp\left(-\frac{1}{\beta_2 \alpha(\mu)}\right)\sum_{m=0}^{\infty}{\frac{(-\tau)^m}{m!}}
\left(\ln\left(\frac{\mu^2}{4Q_s^2}\right)-\frac{1}{\alpha(\mu)\beta_2}\right)^{2m+1} 
\end{eqnarray}
in terms of $\Lambda_{QCD}$
\begin{eqnarray}
 \frac{1}{\alpha(\mu)\beta_2}=\ln{\frac{\mu^2}{\Lambda_{QCD}^2}}
\end{eqnarray}
the non perturbative uncertainty from infrared renormalon found to be,
\begin{eqnarray}
&\sim& r_\perp^2~\Lambda_{QCD}^2~\ln{\frac{\Lambda_{QCD}^2}{4Q_s^2}}~\exp\left({-\tau\ln^2{\frac{\Lambda_{QCD}^2}{4Q_s^2}}}\right)
\end{eqnarray}
Clearly the uncertainty is ${\cal O}\left(\Lambda_{QCD}^2\right)$ instead of ${\cal O}\left(\Lambda_{QCD}^4\right)$ that often taken as characteristic of infrared renormalon. 
Presence of the sudakov factor indicates that the saturation effect 
tend to suppress the IR renormalon effects at small-$x$.

\section{Conclusion}
Infrared renormalon contributions lead to non-perturbative effects similar to those arising from vacuum condensation.
The infrared renormalon contributions presumably contain information about the hadron mass spectrum. 
Corresponding Borel integral is dominated by the singularities closest to the origin. 
In this work we find that small-$x$ effects push the infrared singularities towards the origin and shifts it from $2/\beta_2$ 
to $1/\beta_2$.  
This also makes the  uncertainty proportional to ${\cal O}\left(\Lambda_{QCD}^2\right)$ instead of ${\cal O}\left(\Lambda_{QCD}^4\right)$. 
This very fact however anticipated earlier by Mueller \cite{Mueller:1992xz} 
by citing the fact that perturbation series become ambiguous at 
$b=1/\beta_2$ and was suggested that the $1/Q^2$ should be pursued with vigor. 
In this study we find that ${\cal O}\left(\Lambda_{QCD}^2\right)$ term corresponding to the pole at $1/\beta_2$ exists in unintegrated dipole gluon distribution at small-$x$ as the distribution as a $linear\otimes normal$ distribution in momentum evolved to be a $logarithmic \otimes log~ normal$ distribution at small momentum. 
The result also shows that saturation effects indeed suppress the 
renormalon effect however only through a Sudakov type form factor.


\begin{thebibliography}{100}
\bibitem{Dyson:1952tj} 
  F.~J.~Dyson,
  ``Divergence of perturbation theory in quantum electrodynamics,''
  Phys.\ Rev.\  {\bf 85}, 631 (1952).
  doi:10.1103/PhysRev.85.631

  \bibitem{Lautrup:1977hs}
  B.~E.~Lautrup,
  ``On High Order Estimates in QED,''
  Phys.\ Lett.\  {\bf 69B} (1977) 109.
  doi:10.1016/0370-2693(77)90145-9



\bibitem{Mueller:1984vh} 
  A.~H.~Mueller,
  ``On the Structure of Infrared Renormalons in Physical Processes at High-Energies,''
  Nucl.\ Phys.\ B {\bf 250}, 327 (1985).
  doi:10.1016/0550-3213(85)90485-7





\bibitem{Beneke:1998ui} 
  M.~Beneke,
  ``Renormalons,''
  Phys.\ Rept.\  {\bf 317}, 1 (1999)
  doi:10.1016/S0370-1573(98)00130-6
  [hep-ph/9807443].

 
 
 
 
\bibitem{Zakharov:1992bx} 
  V.~I.~Zakharov,
  ``QCD perturbative expansions in large orders,''
  Nucl.\ Phys.\ B {\bf 385}, 452 (1992).
  doi:10.1016/0550-3213(92)90054-F
  
  
  
\bibitem{Levin:1994di} 
  E.~Levin,
  ``Renormalons at low x,''
  Nucl.\ Phys.\ B {\bf 453}, 303 (1995)
  doi:10.1016/0550-3213(95)00416-P
  [hep-ph/9412345].

  
  



\bibitem{Abir:2018hvk} 
  M.~Siddiqah, N.~Vasim, K.~Banu, R.~Abir and T.~Bhattacharyya,
  ``Unintegrated dipole gluon distribution at small transverse momentum,''
  Phys.\ Rev.\ D {\bf 97}, no. 5, 054009 (2018)
  doi:10.1103/PhysRevD.97.054009
  [arXiv:1801.01637 [hep-ph]].

 \bibitem{Kovchegov_book}
 Y.~V.~Kovchegov and E.~Levin,
  ``Quantum chromodynamics at high energy,''
  Camb.\ Monogr.\ Part.\ Phys.\ Nucl.\ Phys.\ Cosmol.\  {\bf 33}, 1 (2012).
  doi:10.1017/CBO9781139022187
  



\bibitem{Mueller:1992xz} 
  A.~H.~Mueller,
  ``The QCD perturbation series,''
 %
  Talk given at Conference:  QCD: 20 YEARS LATER (Aachen QCD Wkshp.1992, CU-TP-573), proceedings, edited by P.M. Zerwas and H.A. Kastrup, River Edge, N.J., World Scientific, 1993. 2 volumes
p.162-171. 




 \end{thebibliography}
 \end{document}